\documentstyle[12pt,epsfig,subeqn,subeqnarray]{article}

\def\be{\begin{equation}}
\def\ee{\end{equation}}
\def\beq{\begin{eqnarray}}
\def\eeq{\end{eqnarray}}
\def\lsim{\:\raisebox{-0.5ex}{$\stackrel{\textstyle<}{\sim}$}\:}
\def\gsim{\:\raisebox{-0.5ex}{$\stackrel{\textstyle>}{\sim}$}\:} 
\def\GeV{{\rm ~GeV}}

\begin{document}
\begin{flushright}
TIFR/TH/98-08 \\
March 1998
\end{flushright}
\bigskip
\begin{center}
{\Large{\bf Higgs and SUSY Searches at LHC$^*$}} \\[3cm]
{\large D.P. Roy} \\[1cm]
Tata Institute of Fundamental Research, \\ Homi Bhabha Road, 
Mumbai 400 005, India
\end{center}
\bigskip\bigskip

I start with a brief introduction to Higgs mechanism and
supersymmetry.  Then I discuss the theoretical expectations, current
limits and search strategies for Higgs boson(s) at LHC --- first in
the SM and then in the MSSM.  Finally I discuss the signatures and
search strategies for the superparticles.

\vspace{5cm}

\hrule width 5cm

\smallskip

\noindent $^*$ Invited talk at the 5th Workshop on High Energy Physics
Phenomenology (WHEPP-5), Pune, India, 12 - 25 January 1998.

\newpage

As per the Standard Model (SM) the basic constituents of matter are
the quarks and leptons, which interact by the exchange of gauge bosons
-- photon, gluon and the massive $W$ and $Z$ bosons.  By now we have
seen all the quarks and leptons as well as the gauge bosons.  But the
story is not complete yet because of the mass problem.
\bigskip

\noindent {\bf Mass Problem (Higgs Mechanism)}:
\medskip

The question is how to give mass to the weak gauge bosons, $W$ and
$Z$, without breaking gauge symmetry, which is required for a
renormalisable field theory.  In order to appreciate it consider the
weak interaction Lagrangian of a charged scalar field $\phi$; i.e.
\be
{\cal L} = \left(\partial_\mu \phi + ig {\vec\tau \over 2} \vec W_\mu
\phi\right)^\dagger \left(\partial_\mu \phi + ig {\vec\tau \over 2}
\vec W_\mu \phi\right) - \left[\mu^2 \phi^\dagger \phi + \lambda
(\phi^\dagger \phi)^2\right] - {1 \over 4} \vec W_{\mu\nu} \vec
W_{\mu\nu}, 
\ee
where
\be
\vec W_{\mu\nu} = \partial_\mu \vec W_\nu - \partial_\nu \vec W_\mu -
g \vec W_\mu \times \vec W_\nu
\ee
is the field tensor for the weak gauge bosons $\vec W_\mu$.  The
charged and the neutral $W$ bosons form a $SU(2)$ vector, reflecting
the nonabelian nature of this gauge group.  This is responsible for
the last term in (2), which leads to gauge boson
self-interaction.  Correspondingly the gauge transformation on $\vec
W_\mu$ has an extra term, i.e.
\be
\phi \rightarrow e^{i\vec \alpha \cdot \vec \tau} \phi, \ \vec W_\mu
\rightarrow \vec W_\mu - {1 \over g} \partial_\mu \vec\alpha -
\vec\alpha \times \vec W_\mu.
\ee
This ensures gauge invariance of $\vec W_{\mu\nu}$, and hence for the
last term of the Lagrangian, representing gauge kinetic energy.
Evidently the middle term, representing scalar mass and
self-interaction, is invariant under gauge transformation on $\phi$.
Finally the first term, representing scalar kinetic energy and gauge
interaction, can be easily shown to be invariant under the
simultaneous gauge transformations (3).  However the addition of a
mass term 
\be
- M^2 \vec W_\mu \cdot \vec W_\mu,
\ee
would clearly break the gauge invariance of the Lagrangian.  Note
that, in contrast the scalar mass term, $\mu^2 \phi^\dagger \phi$, is
clearly gauge invariant.  This phenomenon is exploited to give mass to
the gauge bosons through back door without breaking the gauge
invariance of the Lagrangian.  This is the celebrated Higgs mechanism
of spontaneous symmetry breaking [1].

One starts with a $SU(2)$ doublet of complex scalar field $\phi$ with
imaginary mass, i.e. $\mu^2 < 0$.  Consequently the minimum of the
scalar potential, $\mu^2 \phi^\dagger \phi + \lambda (\phi^\dagger
\phi)^2$, moves out from the origin to a finite value
\be
v = \sqrt{-\mu^2/\lambda},
\ee
i.e. the field develops a finite vacuum expectation value.  Since the
quantum perturbative expansion is stable only around a local minimum,
one has to translate the field by the constant quantity,
\be
\phi = v + H(x).
\ee
Thus one gets a valid perturbative field theory in terms of the
redefined field $H$.  This represents the physical Higgs boson, while
the 3 other components of the complex doublet field are absorbed to
give mass and hence logitudinal components to the gauge bosons.

Substituting (6) in the first term of the Lagrangian (1) leads to a
mass term for $W$,
\be
M_W = {1\over2} g v.
\ee 
It also leads to a $HWW$ coupling,
\be
{1\over2} g^2 v = g M_W,
\ee
i.e. the Higgs coupling to the gauge bosons is propertional to the
gauge boson mass.  Similarly its couplings to quarks and leptons can
be shown to be propertional to their respective masses, i.e.
\be
h_{\ell,q} = m_{\ell,q}/v = {1\over2} g m_{\ell,q}/M_W.
\ee
Indeed, this is the source of the fermion masses in the SM.  Finally
substituting (6) in the middle term of the Lagrangian leads to a real
mass for the physical Higgs boson,
\be
M_H = v \sqrt{2\lambda} = M_W (2\sqrt{2\lambda}/g).
\ee
Substituting $M_W = 80 \GeV$ and $g = 0.65$ along with a perturbative
limit on the scalar self-coupling $\lambda \lsim 1$, implies that the
Higgs boson mass is bounded by $M_H < 1000 \GeV$.  But the story does
not end here.  Giving mass to the gauge bosons via the higgs mechanism
leads to the so called hierarchy problem.
\bigskip

\noindent {\bf Hierarchy Problem (Supersymmetry)}:
\medskip

\nobreak
The problem is how to peg down the Higgs scalar in the desired mass
range of a few hundered $\GeV$.  This is because the scalar masses are
known to have quadratically divergent quantum corrections from
radiative loops involving e.g. quarks or leptons.  These would push
the output scalar mass to the cut-off scale of the SM, i.e. the GUT
scale $(10^{16} \GeV)$ or the Planck scale $(10^{19} \GeV)$.  The
desired mass range of $\sim 10^2 \GeV$ is clearly tiny compared to
these scales.  This is the so called hierarchy problem.  The
underlying reason for the quadratic divergence is that the scalar
masses are not protected by any 
symmetry unlike the fermion and the gauge boson masses, which are
protected by chiral symmetry and gauge symmetry.  Of course it was
this very property of the scalar mass that was exploited to give
masses to the fermions and gauge bosons in the first place.  Therefore
it can not be simply wished away.

The most attractive solution to this problem is provided by
supersymmetry (SUSY), a symmetry between fermions and bosons [2].  It
predicts the quarks and leptons to have scalar superpartners called
squarks and sleptons $(\tilde q, \tilde \ell)$, and the gauge bosons
to have fermionic superpartners called gauginos $(\tilde g,\tilde \gamma,
\tilde W, \tilde Z)$.  In the minimal supersymmetric extension of the
standard model (MSSM) one needs two Higgs doublets $H_{1,2}$, with
opposite hypercharge $Y = \pm 1$, to give masses to the up and down
type quarks.  The corresponding fermionic superpartners are called
Higgsinos $(\tilde H_{1,2})$.  The opposite hypercharge of these two
sets of fermions ensures anomaly cancellation. 

SUSY ensures that the quadratically divergent quantum corrections from
quark and lepton loops are cancelled by the contributions from the
corresponding squark and slepton loops.  Thus the Higgs masses can be
kept in the desired range of $\sim 10^2 \GeV$.  However this implies
two important constraints on SUSY breaking. 

\begin{enumerate}
\item[{i)}] SUSY can be broken in masses but not in couplings (soft
breaking), so that the co-efficients of the cancelling contributions
remain equal and opposite.
\item[{ii)}] The size of SUSY breaking in masses is $\sim 10^2 \GeV$,
so that the size of the remainder remains within this range.  Thus the
superpartners of the SM particles are also expected to lie in the mass
range of $\sim 10^2 \GeV$, going upto $1000 \GeV$.
\end{enumerate}
\bigskip

\noindent {\bf SM Higgs Boson}: {\bf Theoretical Constraints
\& Search Strategy}
\medskip

\nobreak
The Higgs self coupling $\lambda$ is ultra-violet divergent.  It
evolves according to the renormalisation group equation (RGE)
\be
{d\lambda \over d \ell n(\mu/M_W)} = {3\lambda^2 \over 2\pi^2}.
\ee
It can be easily solved to give
\be
\lambda (\mu) = {1 \over 1/\lambda (M_W) - (3/2\pi^2) \ell
n(\mu/M_W)},
\ee
which has a Landau pole at 
\beq
\mu_\infty &=& M_W e^{2\pi^2/3\lambda (M_W)}, \nonumber \\[2mm]
\lambda (M_W) &=& {g^2 \over 8} {M^2_H \over M^2_W}.
\eeq
Thus the larger the starting value $\lambda (M_W)$, the sooner will the
coupling diverge.  This is illustrated in Fig. 1.  Evidently the
theory is valid only upto a cut-off scale $\wedge = \mu_\infty$.
Requiring the theory to be valid at all energies, $\wedge \rightarrow
\infty$, would imply $\lambda (M_W) \rightarrow 0$; i.e. the only good
$\lambda \phi^4$ theory is a trivial theory.  Surely we do not want
that.  But if we want the theory to be valid upto the Planck scale or
GUT scale, we must have a relatively small $\lambda (M_W)$, which
corresponds to a small $M_H \lsim 200 \GeV$.  If on the other hand we
assume it to be valid only upto the TeV scale, then we can have a
larger $\lambda (M_W)$, corresponding to a relatively large $M_H \lsim
600 \GeV$.  This is the so-called triviality bound [3].  If $M_h$ is
significantly larger than $600 \GeV$, then the range of validity of
the theory is limited to $\wedge < 2 M_H$.  This would correspond to a
composite Higgs scenario, e.g. technicolour models.

Fig. 2 shows the triviality bound on the Higgs mass against the
cut-off scale $\wedge$ of the theory [4].  It also shows a lower bound
on the Higgs mass, which comes from a negative contribution to the RGE
(11) from the top Yukawa coupling, i.e.
\be
{d\lambda \over d \ell n (\mu/M_W)} = {3 \over 2\pi^2} (\lambda^2 +
\lambda h^2_t - h^4_t).
\ee
The Yukawa coupling being ultra-violet divergent turns $\lambda$
negative at a high energy scale; and the smaller the starting value of
$\lambda$ (or equivalently $M_H$) the sooner will it become negative.
A negative $\lambda$ coupling has the undesirable feature of an
unstable vacuum (eq. 5).  Thus one can define a cut-off scale $\wedge$
for the theory, where this change of sign occurs.  The lower curve of
Fig. 2 shows the lower bound on $M_H$ as a function of the cut-off
scale $\wedge$ including the theoretical uncertainty [5].  We see from
this figure that the longer the range of validity of the theory, the
stronger will be the upper and lower bounds on $M_H$.  Thus assuming
no new physics upto the GUT or Planck scale (the desert scenario)
would constrain the SM Higgs mass to lie in the range
\be
M_H = 130 - 190 \GeV.
\ee
However the lower bound becomes invalid once we have more than one
Higgs doublet, since the unique relation between the top mass and
Yukawa coupling (9) will no longer hold.  In particular, one expects
an upper bound of $\sim 130 \GeV$ for the lightest Higgs boson mass in
MSSM in stead of a lower bound, as we shall see below.  Since one
needs SUSY or some other form of new physics to stabilize the Higgs
mass, the above vacuum stability bound may have limited significance.
Nonetheless it is interesting to note that the predicted range of the
SM Higgs boson mass (15) agrees favourably with the indirect estimate
of this quantity from the precision measurement of electro-weak
parameters at LEP/SLD [6], i.e.
\be
M_H = 115^{+116}_{-66} \GeV \ \left(< 420 \GeV \ {\rm at} \ 95\% \
{\rm CL}\right).
\ee
It should be added however that there is a lingering discrepancy
between the LEP and the SLD values of $\sin^2 \theta_W$, which could
affect the central value and the 95\% CL limit of $M_H$ appreciably.
Thus all one can say at the moment is that these indirect estimates
are consistent with a relatively light Higgs boson.

The search strategy for Higgs boson is based on its preferential
coupling to the heavy quarks and gauge bosons as seen from (8,9).  The
LEP-I search was based on the so called Bjorken process
\be
e^+e^- \rightarrow Z \rightarrow HZ^\star \rightarrow \bar bb (\ell^+
\ell^-, \nu\nu, \bar qq),
\ee
while the LEP-II search is based on the associated process with $Z$
and $Z^\star$ interchanged.  The current LEP-II limit from the
preliminary ALEPH data at $183 \GeV$ is [7]
\be
M_H > 88.6 \GeV.
\ee
The forthcoming runs at $192 - 200 \GeV$ are expected to extend the
search upto
\be
M_H = 95 - 100 \GeV.
\ee

Thus the Higgs mass range of interest to LHC is $M_H \gsim 90 \GeV$.
Fig. 3 shows the total decay width of the Higgs boson over this range
along with the branching ratios for the important decay channels [8].
It is clear from this figure that the mass range can be divided into
two parts -- a) $M_H < 2 M_W (90 - 160 \GeV)$ and b) $M_H > 2M_W (160
- 1000 \GeV)$.

The first part is the so called intermediate mass region, where the
Higgs width is expected to be only a few MeV.  The dominant decay mode
is $H \rightarrow \bar bb$.  This has unfortunately a huge QCD
background, which is $\sim 1000$ times larger than the signal.  By far
the cleanest channel is $\gamma\gamma$, where the continuum background
is a 2nd order $EW$ process.  However, it suffers from a small
branching ratio 
\be
B (H \rightarrow \gamma\gamma) \sim 1/1000,
\ee
since it is a higher order process, induced by the top quark loop.  So
one needs a very high jet/$\gamma$ rejection factor $\gsim 10^8$.
Besides the continuum background being propertional to $\Delta
M_{\gamma\gamma}$, one needs a high resolution,
\be
\Delta M_{\gamma\gamma} \lsim 1 \GeV \ {\rm i.e.} \ \lsim 1\% \ {\rm
of} \ M_H.
\ee
This requires fine $EM$ calorimetry, capable of measuring the $\gamma$
energy and direction to $1\%$ accuracy.  In this respect CMS is
expected to do better than ATLAS.  The projected Higgs mass reach of
the two detectors via this channel are $M_H = 90 - 140 \GeV$ (CMS) and
$110 - 140 \GeV$ (ATLAS) at the high luminosity run of LHC ($100
fb^{-1}$). 

One can get a feel for the size of the signal from the Higgs
production cross-sections shown in Fig. 4.  The relevant production
processes are 
\be
gg \ {\buildrel {\bar t^\star t^\star} \over \longrightarrow} \ H,
\ee
\be
qq \ {\buildrel {W^\star W^\star} \over \longrightarrow} \ H q q,
\ee
\be
q\bar q' \ {\buildrel {W^\star} \over \rightarrow} \ HW,
\ee
\be
gg, q\bar q \rightarrow Ht \bar t (Hb \bar b).
\ee
The largest cross-section, coming from gluon-gluon fusion via the top
quark loop (22), is of the order of $10 pb$.  Thus the expected size
of the $H \rightarrow \gamma\gamma$ signal is $\sim 10 fb$,
corresponding to $\sim 10^3$ events.  The estimated continuum
background is $\sim 10^4$ events, which can of course be subtracted
out.  Thus the significance of the signal is given by its relative
size with respect to the statistical uncertainty in the background,
i.e.
\be
S/\sqrt{B} \simeq 10.
\ee
By far the cleanest signal is provided by the associated Bjorken
process (24), with a cross-section of $\sim 1 fb$ in the $H
\rightarrow \gamma\gamma$ channel.  Combining this with the $BR$ of
$2/9$ for $W \rightarrow \ell \nu$ implies a signal of $20 - 30$
events in the $\ell + \gamma \gamma$ channel.  While the signal size
is admittedly small, the estimated background is only $\sim 10$
events.  Thus the $S/\sqrt{B}$ ratio is again $\sim 10$.  Detailed
signal and background simulations for these channels can be found in
[9].  The result is summarised in Fig. 5.  It shows that one expects a
$5\sigma$ signal upto a Higgs mass of $150 \GeV$ for an integrated
Luminosity of $30 fb^{-1}$.  This corresponds to the low luminosity
run of LHC over the first 3 years.  It may be noted that the dominant
decay channel, $H \rightarrow b\bar b$, can be important for Higgs
search below $100 \GeV$.  It comes from the associated Bjorken process (24),
where the leptonic decay of the accompanying $W(Z)$ helps to reduce
the background.  However this region should be already covered by
LEP-II. 

The most promising Higgs decay channel is
\be
H \rightarrow ZZ \rightarrow \ell^+ \ell^- \ell^+ \ell^-,
\ee
since reconstruction of the $\ell^+ \ell^-$ invariant masses makes it
practically background free.  Thus it provides the most important
Higgs signal right from the subthresold region of $M_H = 140 \GeV$
upto $600 \GeV$ (see Fig. 3).  Note however a sharp dip in the $ZZ$
branching ratio at $M_H = 160 - 170 \GeV$ due to the opening of the
$WW$ channel.  The most important Higgs signal in this dip region is
expected to come from [10]
\be
H \rightarrow WW \rightarrow \ell^+ \nu \ell^- \bar\nu.
\ee
However in general this channel suffers from a much larger background
for two reasons -- i) it is not possible to reconstruct the $W$ masses
because of the two neutrinos and ii) there is a large $WW$ background
from $t\bar t$ decay.

For large Higgs mass, $M_H = 600 - 1000 \GeV$, the 4-lepton signal
(27) becomes too small in size.  In this case the decay channels 
\be
H \rightarrow WW \rightarrow \ell \nu q \bar q', \ \ H \rightarrow ZZ
\rightarrow \ell^+ \ell^- q \bar q,
\ee
are expected to provide more favourable signals.  The biggest
background comes from single $W(Z)$ production along with QCD jets.
However, one can exploit the fact that a large part of the signal
cross-section in this case comes from $WW$ fusion (23), which is
accompanied by two forward (large-rapidity) jets.  One can use the
double forward jet tagging to effectively control the background.
Indeed simulation studies by the CMS and ATLAS collaborations show
that using this strategy one can extend the Higgs search right upto
$1000 \GeV$ [9].
\bigskip

\noindent {\bf MSSM Higgs Bosons: Theoretical Constraints \&
Search Strategy}
\medskip

\nobreak
As mentioned earlier, the MSSM contains two Higgs doublets, which
correspond to 8 independent states.  After 3 of them are absorbed by the
$W$ and $Z$ bosons, one is left with 5 physical states: two neutral
scalars $h^0$ and $H^0$, a pseudoscalar $A^0$, and a pair of charged
Higgs scalars $H^\pm$.  At the tree-level their masses and couplings
are determined by only two parameters -- the ratio of the two vacuum
expectation values, $\tan\beta$, and one of the scalar masses, usually
taken to be $M_A$.  However, the neutral scalars get a large radiative
correction from the top quark loop along with the top squark (stop)
loop.  To a good approximation this is given by [11]
\be
\epsilon = {3g^2 m^4_t \over 8\pi^2 M^2_W} \ell n\left({M^2_{\tilde
t} \over m^2_t}\right),
\ee
plus an additional contribution from the $\tilde t_{L,R}$ mixing,
\be
\epsilon_{\rm mix} = {3g^2 m^4_t \over 8\pi^2 M^2_W} {A^2_t \over
M^2_{\tilde t}} \left(1 - {A^2_t \over 12 M^2_{\tilde t}}\right) \leq
{9g^2 m^4_t \over 8\pi^2 M^2_W}.
\ee
Thus while the size of $\epsilon_{\rm mix}$ depends on the trilinear
SUSY breaking parameter $A_t$, it has a maximum value independent of
$A_t$.  As expected the radiative corrections vanish in the exact SUSY
limit.  One can estimate the rough magnitude of these corrections
assuming a SUSY breaking scale of $M_{\tilde t} = 1$ TeV.  The leading
log QCD corrections can be taken into account by using the running
mass of top at the appropriate energy scale [11]; i.e. $m_t (\sqrt{m_t
M_{\tilde t}}) \simeq 157 \GeV$ in (30) and $m_t (M_{\tilde t}) \simeq
150 \GeV$ in (31) instead of the top pole mass of $175 \GeV$.  One
can easily check that the resulting size of the radiative corrections are 
\be
\epsilon \sim M^2_W \ \ {\rm and} \ \ 0 < \epsilon_{\rm mix} \lsim
M^2_W. 
\ee

The neutral scalar masses are obtained by diagonalising the
mass-squared matrix
\be
\left(\matrix{M^2_A \sin^2 \beta + M^2_Z \cos^2 \beta & -(M^2_A +
M^2_Z) \sin \beta \cos \beta \cr & \cr -(M^2_A + M^2_Z) \sin \beta
\cos \beta & M^2_A \cos^2 \beta + M^2_Z \sin^2 \beta +
\epsilon'}\right)
\ee
with $\epsilon' = (\epsilon + \epsilon_{\rm mix})/\sin^2 \beta$.  Thus
\beq
M^2_h &=& {1\over2} \Bigg[M^2_A + M^2_Z + \epsilon' - \Big\{(M^2_A +
M^2_Z + \epsilon')^2 - 4M^2_A M^2_Z \cos^2 \beta \nonumber \\[2mm]
& & ~~~~ - 4\epsilon' (M^2_A \sin^2 \beta + M^2_Z \cos^2
\beta)\Big\}^{1/2}\Bigg] \nonumber \\[2mm]
M^2_H &=& M^2_A + M^2_Z + \epsilon' - M^2_h \nonumber \\[2mm]
M^2_{H^\pm} &=& M^2_A + M^2_W
\eeq
where $h$ denotes the lighter neutral scalar [12].  One can easily
check that its mass has an upper bound for $M_A \gg M_Z$, i.e.
\be
M^2_h \longrightarrow M^2_Z \cos^2 2\beta + \epsilon + \epsilon_{\rm
mix}, 
\ee
while $M^2_H$, $M^2_{H^\pm} \rightarrow M^2_A$.  Thus the MSSM
contains at least one light Higgs boson $h$, whose tree-level mass
limit $M_h < M_Z$, goes upto $130 - 140 \GeV$ after including the
radiative corrections.  Fig. 6 shows the masses of the MSSM Higgs
bosons against $M_A$ for two representative values of $\tan \beta = 1.5$
and $30$.  The predictions without stop mixing and with maximal mixing
are shown in separate plots.  Note that the $h$ mass limit is
particularly strong in the low $\tan\beta$ $(\sim 1)$ region,
i.e. $M_h < 80 - 100 \GeV$ depending on the size of stop mixng
parameter $A_t$.  Consequently the low $\tan\beta$ region is
particularly succeptible to the ongoing Higgs search at LEP-II as we
shall see below.

Let us consider now the couplings of the MSSM Higgs bosons.  A
convenient parameter for this purpose is the mixing angle $\alpha$
between the neutral scalars, i.e.
\be
\tan 2\alpha = \tan 2\beta {M^2_A + M^2_Z \over M^2_A - M^2_Z +
\epsilon'/\cos 2\beta}, -\pi/2 < \alpha < 0.
\ee
Note that
\be
\alpha \ {\buildrel {M_A \gg M_Z} \over \longrightarrow} \ \beta -
\pi/2. 
\ee
\begin{enumerate}
\item[{}] Table-I.  Important couplings of the MSSM Higgs bosons $h$,
$H$ and $A$ relative to those of the SM Higgs boson
\end{enumerate}
\[
\begin{tabular}{|c|c|c|c|c|}
\hline
&&&& \\
Channel & $H_{\rm SM}$ & $h$ & $H$ & $A$ \\
&&&& \\
\hline
&&&& \\
$\bar bb(\tau^+\tau^-)$ & $\displaystyle{gm_b \over 2M_W} (m_\tau)$ & $\sin
\alpha/\cos \beta$ & $\cos \alpha/\cos \beta$ & $\tan \beta$ \\
 & & $\rightarrow 1$ & $\tan \beta$ & '' \\
&&&& \\
\hline
&&&& \\
$\bar tt$ & $\displaystyle g{m_t \over 2M_W}$ & $\cos\alpha/\sin\beta$ &
$\sin\alpha/\sin\beta$ & $\cot \beta$ \\
& & $\rightarrow 1$ & $\cot\beta$ & '' \\
&&&& \\
\hline
&&&& \\
$WW (ZZ)$ & $g M_W (M_Z)$ & $\sin (\beta - \alpha)$ & $\cos (\beta -
\alpha)$ & $0$ \\
& & $\rightarrow 1$ & $0$ & '' \\
&&&& \\
\hline
\end{tabular}
\]

\noindent Table-I shows the important couplings of the neutral Higgs bosons
relative to those of the SM Higgs boson.  The limiting values of these
couplings at large $M_A$ are indicated by arrows.  The corresponding
couplings of the charged Higgs boson, which has no SM analogue, are 
\beq
H^+ \bar t b &:& {g \over \sqrt{2}M_W} (m_t \cot \beta + m_b \tan
\beta), \ H^+ \tau \nu : {g \over \sqrt{2}M_W} m_\tau \tan \beta,
\nonumber \\[2mm]
H^+ W^- Z &:& 0.
\eeq
Note that the top Yukawa coupling is ultraviolet divergent.  Assuming
it to lie within the perturbation theory limit all the way upto the
GUT scale implies 
\be
1 < \tan \beta < m_t/m_b,
\ee
which is therefore the favoured range of $\tan\beta$.  However, it
assumes no new physics beyond the MSSM upto the GUT scale, which is a
stronger assumption than MSSM itself.  Nontheless we shall concentrate
in this range.

Coming back to the neutral Higgs couplings of Table-I, we see that in
the large $M_A$ limit the light Higgs boson $(h)$ couplings approach
the SM values.  The other Higgs bosons are not only heavy, but their
most important couplings are also suppressed.  This is the so
called decoupling limit, where the MSSM Higgs sector is
phenomenologically indistinguishable from the SM.  It follows
therefore that the Higgs search stategy for $M_A \gg M_Z$ should be
the same as the SM case, i.e. via 
\be
h \rightarrow \gamma\gamma.
\ee

At lower $M_A$, several of the MSSM Higgs bosons become light.
Unfortunately their couplings to the most important channels, $\bar
tt$ and $WW/ZZ$, are suppressed relative to the SM Higgs boson [12].
Thus their most important production cross-sections as well as their
decay BRs into the $\gamma\gamma$ channel are suppressed relative to
the SM case.  Consequently the Higgs detection in this region is very
hard and it calls for multiprong stategy from the three sides in
the $M_A - \tan\beta$ plane (Fig. 7): (a) Low $M_A$, (b) High
$\tan\beta$ and (c) Low $\tan\beta$.

\begin{enumerate}
\item[{(a)}] Low $M_A (\lsim M_Z)$ -- In this case $M_{H^\pm} < M_t$;
and the best strategy is to search for $H^\pm$ in top quark decay,
i.e.
\be
t \rightarrow bH^+, \ H^+ \rightarrow \tau \nu,
\ee
via preferential top decay into the $\tau$ channel as well as the
opposite polarization of $\tau$ wrt the SM decay $(W \rightarrow \tau
\nu)$ [14].

\item[{(b)}] High $\tan \beta (\sim m_t/m_b)$ -- It is clear from
Table-I that in this case the best production and decay channels are 
\be
gg \rightarrow b\bar b (h,H,A) \rightarrow b\bar b \tau^+ \tau^-.
\ee

\item[{(c)}] Low $\tan\beta (\sim 1)$ -- As mentioned earlier the LEP-II
search via the associated Bjorken process becomes very effective in
this case.  Indeed one can see from (36) that in this case (37) holds
even for relatively low $M_A$, so that the $hZZ$ coupling is very
close to the SM case.  Thus the present LEP-II limit (18) as well as
the discovery limit (19) are equally valid for $M_h$ in the low
$\tan\beta$ region.  As one can see from Fig. 6, the former rules out
the $\tan\beta \lsim 1.5$ region for the no stop-mixing case while the
latter will rule it out even for the maximal stop mixing case.  It may
be added here that a modest part of the large $\tan\beta$ region seems
to be ruled out as well by the Tevatron data via (41) [15] and (42) [16].
\end{enumerate}

Fig. 7 summarises the MSSM Higgs discovery limits of the CMS detector
at LHC via the three processes (40-42) along with the LEP-II limit.
Note that there is a significant hole in the $M_A - \tan\beta$ plane
that is left out even after combining all the 4 limits.  Moreover the
LEP-II limit will go down to a lower range of $\tan\beta$ when stop
mixing effect is taken into account (Fig. 6).  This will enlarge the
size of the hole further.  Finally, Fig. 8 shows that it would be
possible to close the hole if one combines the CMS and the ATLAS data
collected over an integrated luminosity of $300 fb^{-1}$ [17].  This
corresponds to 3 years of high luminosity run of LHC; and illustrates
the challenge involved in the search for the MSSM Higgs bosons.  Note
that even in this case there is a large region where one would see
only one Higgs boson $(h)$ with SM like couplings and hence not be
able to distinguish the SUSY from the SM Higgs sector.  Fortunately it
will be possible and in fact much easier to probe SUSY directly via
superparticle search as we see below.
\bigskip

\noindent {\bf Superparticles: Signature \& Search Strategy}
\medskip

\nobreak
I shall concentrate on the standard SUSY model, where the
superparticle signature is based on $R$-parity conservation.  Let me
start therefore with a brief discussion of $R$-parity.  The presence
of scalar quarks in SUSY can lead to baryon and lepton number
violating interactions of the type $ud \rightarrow \bar{\tilde s}$ and
$\bar {\tilde s} \rightarrow e^+ \bar u$, i.e.
\be
u d \ {\buildrel {\bar{\tilde s}} \over \longrightarrow} \ e^+ \bar u.
\ee
Moreover adding a spectator $u$ quark to both sides one sees that this
can lead to a catastrophic proton decay, i.e. 
\be
p (uud) \ {\buildrel {\bar{\tilde s}} \over \longrightarrow} \ e^+ \pi^0
(\bar u u).
\ee
Since the superparticle masses are assumed to be of the order $M_W$
for solving the hierarchy problem, this would imply a proton life time
similar to the typical weak decay time of $\sim 10^{-8} {\rm sec}$!
The best way to avoid this catastrophic proton decay is via $R$-parity
conservation, where
\be
R = (-1)^{3B+L+2S}
\ee
is defined to be $+1$ for the SM particles and $-1$ for their
superpartners, since they differ by $1/2$ unit of spin $S$.  It
automatically ensures $L$ and $B$ conservation by preventing single
emission (absorption) of superparticle.

Thus $R$-conservation implies that (i) superparticles are produced in
pair and (ii) the lightest superparticle (LSP) is stable.
Astrophysical evidences against such a stable particle carrying colour
or electric charge imply that the LSP is either sneutrino $\tilde \nu$
or photino $\tilde\gamma$ (or in general the lightest neutralino).
The latter alternative is favoured by the present SUSY models.  In
either case the LSP is expected to have only weak interaction with
ordinary matter like the neutrino, since e.g.
\be
\tilde\gamma q \ {\buildrel {\tilde q} \over \longrightarrow} \ q
\tilde \gamma \ \ {\rm and} \ \ \nu q \ {\buildrel W \over
\longrightarrow} \ e q'
\ee
have both electroweak couplings and $M_{\tilde q} \sim M_W$.  This
makes the LSP an ideal candidate for the cold dark matter.  It also
implies that the LSP would leave the normal detectors without a trace
like the neutrino.  The resulting imbalance in the visible momentum
constitutes the canonical missing transverse-momentum $(p\!\!\!/_T)$
signature for superparticle production at hadron colliders.  It is
also called the missing transverse-energy $(E\!\!\!/_T)$ as it is
often measured as a vector sum of the calorimetric energy deposits in
the transverse plane.

The main processes of superparticle production at LHC are the QCD
processes of quark-antiquark and gluon-gluon fusion [18]
\be
q\bar q, gg \longrightarrow \tilde q \bar{\tilde q} (\tilde g \tilde
g).
\ee
The NLO corrections can increase these cross-sections by $15-20\%$
[19].  The simplest decay processes for the produced squarks and
gluinos are
\be
\tilde q \rightarrow q \tilde \gamma, \ \tilde g \rightarrow q\bar q
\tilde \gamma.
\ee
Convoluting these with the pair production cross-sections (47) gives
the simplest jets + $p\!\!\!/_T$ signature for squark/gluino
production, which were adequate for the early searches for relatively
light squarks and gluinos.  However, over the mass range of current
interest $(\geq 100 \GeV)$ the cascade decays of squark and gluino
into the LSP via the heavier chargino/neutralino states are expected
to dominate over the direct decays (48).  This is both good news and
bad news.  On the one hand the cascade decay degrades the missing-$p_T$
of the canonical jets $+ p\!\!\!/_T$ signature.  But on the other hand
it gives a new multilepton signature via the leptonic decays of these
chargino/neutralino states.  It may be noted here that one gets a mass
limit of
\be
M_{\tilde q,\tilde g} > 180 \GeV
\ee
from the Tevatron data using either of the two signatures [20].

The cascade decay is described in terms of the $SU(2) \times U(1)$
gauginos $\tilde W^{\pm,0}, \tilde B^0$ along with the Higgsinos
$\tilde H^\pm$, $\tilde H^0_1$ and $\tilde H^0_2$.  The $\tilde B$ and
$\tilde W$ masses are denoted by $M_1$ and $M_2$ respectively while
the Higgsino masses are functions of the supersymmetric Higgsino mass
parameter $\mu$ and $\tan\beta$.  The charged and the neutral gauginos
will mix with the corresponding Higgsinos to give the physical
chargino $\chi^\pm_{1,2}$ and neutralino $\chi^0_{1,2,3,4}$ states.
Their masses and compositions can be found by diagonalising the
corresponding mass matrices, i.e.
\[
M_C = \left(\matrix{M_2 & \sqrt{2} M_W \sin \beta \cr & \cr \sqrt{2}
M_W \cos \beta & \mu}\right),
\]
\bigskip
\be
M_N = \left(\matrix{M_1 & 0 & -M_Z \sin \theta_W \cos \beta & M_Z \sin
\theta_W \sin \beta \cr & & & \cr 0 & M_2 & M_Z \cos \theta_W \cos
\beta & -M_Z \cos \theta_W \sin \beta \cr & & & \cr -M_Z \sin \theta_W
\cos \beta & M_Z \cos \theta_W \cos \beta & 0 & -\mu \cr & & & \cr M_Z
\sin \theta_W \sin \beta & -M_Z \cos \theta_W \sin \beta & -\mu &
0}\right).
\ee

Let me try to present a simplified picture.  Assuming unification of
the $SU(3) \times SU(2) \times U(1)$ gaugino masses at the GUT scale
the RGE relates the corresponding masses at the low energy scale
$(\sim 10^2 \GeV)$ to the respective gauge couplings.  Thus
\beq
M_2 &=& (g^2/g^2_S) M_{\tilde g} \simeq 0.3 M_{\tilde g} \nonumber
\\[2mm]
M_1 &=& (5\tan^2 \theta_W/3) M_2 \simeq 0.5 M_2.
\eeq
Moreover the SUGRA assumptions of a common scalar mass at the GUT
scale along with the radiative breaking of the $EW$ symmetry (5),
imply
\be
\mu \gg M_2.
\ee
It is clear from (50,51,52) that the lighter chargino and neutralino
states are expected to be dominated by gaugino components.  In
particular
\beq
\chi^\pm_{1,2} &\simeq& \tilde W^\pm, \tilde H^\pm \nonumber \\[2mm]
\chi^0_{1,2,\cdots} &\simeq& \tilde B, \tilde W^0, \cdots
\eeq

With the above systematics one can understand the essential features
of cascade decay.  For illustration I shall briefly discuss cascade
decay of gluino for two representative gluino mass regions of interest
to LHC.

\begin{enumerate}
\item[{i)}] $M_{\tilde g} \simeq 300 \GeV$: In this case the gluino
decays into the light quarks
\be
g \rightarrow \bar q q \left[\tilde B (.2), \tilde W^0 (.3). \tilde
W^\pm (.5)\right], \ q \neq t,
\ee
which have negligible Yukawa couplings.  Thus the decay branching
ratios are propertional to the squares of the respective gauge
couplings as indicated in parantheses.  Because of the smaller $U(1)$
gauge coupling relative to the $SU(2)$, the direct decay into the LSP
$(\tilde B)$ is small compared to cascade decay via the heavier
($\tilde W$ dominated) chargino and neutralino states.  The latter
decay into the LSP via real or virtual $W/Z$ emission,
\be
\tilde W_0 \rightarrow Z \tilde B \rightarrow \ell^+ \ell^- \tilde B
(.06), \ \tilde W^\pm \rightarrow W\tilde B \rightarrow \ell^\pm \nu
\tilde B (.2),
\ee
whose leptonic branching ratios are indicated in parantheses.  From
(54) and (55) one can easily calculate the branching ratios of
dilepton and trilepton states resulting from the decay of a gluino
pair.  In particular the dilepton final state via the charginos has a branching ratio of
$1\%$.  Then the Majorana nature of $\tilde g$ implies a distinctive
like sign dilepton (LSD) signal with a BR of $\sim 1/2\%$.

\item[{ii)}] $\tilde M_{\tilde g} \gsim 500 \GeV$: In this case the
large top Yukawa coupling implies a significant decay rate via
\be
\tilde g \rightarrow t \bar b \tilde H^-,
\ee
where both $t$ and $\tilde H^-$ can contribute to the leptonic final
state via
\be
t \rightarrow bW^+ \rightarrow b \ell^+ \nu (.2), \ \tilde H^-
\rightarrow W^- \tilde B \rightarrow \ell^- \nu \tilde B (.2).
\ee
Consequently the BR of the LSD signal from the decay of the gluino
pair is expected to go up to $3-4\%$.
\end{enumerate}

Fig. 9 shows the expected LSD signal from gluino pair production at
LHC for $M_{\tilde g} = 300$ and $800 \GeV$ along with the background
[21].  The latter comes from $\bar tt$ via cascade decay (long dashed)
or charge misidentification (dots).  Note that the signal is
accompanied by a much larger $p\!\!\!/_T$ compared to the background
because of the LSPs.  This can be used to effectively suppress the
background while retaining about $1/2$ of the signal.  Consequently
one can search for a gluino upto at least $800 \GeV$ at the low
luminosity $(10 fb^{-1})$ run of LHC, going upto $1200 \GeV$ at the
high luminosity $(100 fb^{-1})$.

Fig. 10 shows the size of the canonical $p\!\!\!/_T +$ jets signal
against gluino mass for two cases -- $M_{\tilde g} \ll M_{\tilde q}$
(triangles) and $M_{\tilde g} \simeq M_{\tilde q}$ (squares) [22].
The background line shown also corresponds to $5\sqrt{B}$ for the
LHC luminosity of $10 fb^{-1}$.  Thus one expects a $5\sigma$
discovery limit of at least upto $M_{\tilde g} = 1300 \GeV$ from this
signal\footnote{This signal may be hard to observe at the high
luminosity run due to event pileup.}.  Finally, Fig. 11 shows the CMS
simulation for the $5\sigma$ discovery limits from the various
leptonic channels in the plane of $m_0 - m_{1/2}$, the common scalar
and gaugino masses at the GUT scale.  The corresponding squark and
gluino mass contours are also shown.  As we see from this figure, it
will be possible to extend the squark and gluino searches at LHC well
into the TeV region.  One should either see these superparticles or
rule out SUSY at least as a solution to the hierarchy problem of the
SM. 

I am thankful to Dr. D. Denegri of the CMS collaboration for providing
figures 7 and 11.  I also thank R.S. Pawar and S.K. Vempathi for help
in plotting figures 1 and 6 and S. Datta for embedding the figure files. 

\newpage

\noindent {\bf References}:
\medskip

\begin{enumerate}
\item[{1.}] For a review see e.g. J.F. Gunion, H.E. Haber, G. Kane and
S. Dawson, The Higgs Hunters' Guide (Addison-Wesley, Reading, MA,
1990). 
\item[{2.}] For a review see e.g. H.E. Haber and G. Kane,
Phys. Rep. 117, 75 (1985). 
\item[{3.}] L. Maiani, G. Parisi and R. Petronzio, Nucl. Phys. B136,
115 (1978); N. Cabbibo, L. Maiani, G. Parisi and R. Petronzio,
Nucl. Phys. B158, 295 (1979); M. Lindner, Z. Phys. C31, 295 (1986).
\item[{4.}] K. Riesselmann, hep-ph/9711456 (to appear in Proc. 35th
Intl. School of Subnuclear Physics, Erice, 1997).  For the theoretical
uncertainty in the upperbound see T. Hambye and K. Riesselmann,
Phys. Rev. D55, 7255 (1997). 
\item[{5.}] G. Altarelli and G. Isidori, Phys. Lett. B337, 141 (1994);
J.A. Casas, J.R. Espinosa and M. Quiros, Phys. Lett. B342, 171 (1995)
and B382, 374 (1996). 
\item[{6.}] LEP Electroweak Working Group and SLD Heavy Flavour Group,
CERN-PPE/97-154 (2 Dec. 1997).
\item[{7.}] P. Janot, Invited talk at the Indo-French Workshop on
Supersymmetry and Unification, TIFR, Mumbai (17-21 Dec. 1997). 
\item[{8.}] M. Spira, hep-ph/9705337.
\item[{9.}] CMS Technical Proposal, CERN/LHCC/94-38 (1994); ATLAS
Technical Proposal, CERN/LHCC/94-43 (1994).
\item[{10.}] M. Dittmar and H. Dreiner, Phys. Rev. D55, 167 (1997). 
\item[{11.}] For a recent review of radiative correction along with
reference to the earlier works see H.E. Haber, R. Hempfling and
A.H. Hoang, Z. Phys. C75, 539 (1997).
\item[{12.}] See e.g. A. Djouadi, J. Kalinowski and P.M. Zerwas,
Z. Phys. C70, 435 (1996).
\item[{13.}] Simulation study by the CMS collaboration. 
\item[{14.}] D.P. Roy, Phys. Lett. 277B, 183 (1992) and
Phys. Lett. 283B, 403 (1992); S. Raychaudhuri and D.P. Roy,
Phys. Rev. D53, 4902 (1996).  
\item[{15.}] M. Guchait and D.P. Roy, Phys. Rev. D55, 7263 (1997); CDF
Collaboration: F. Abe et al., Phys. Rev. Lett. 79, 357 (1997).
\item[{16.}] M. Drees, M. Guchait and P. Roy, hep-ph/9801229.
\item[{17.}] E. Richter-Was et al., CERN-TH-96-111 (1996). 
\item[{18.}] G.L. Kane and J.P. Leville, Phys. Lett. B112, 227 (1982);
P.R. Harrison and C.H. Llewellyn-Smith, Nucl. Phys. B213, 223 (1983)
[Err. Nucl. Phys. B223, 542 (1983)]; E. Reya and D.P. Roy,
Phys. Rev. D32, 645 (1985).
\item[{19.}] W. Beenakker, R. Hopker, M. Spira and P. Zerwas,
Nucl. Phys. B492, 51 (1997); M. Kramer, T. Plehn, M. Spira and
P. Zerwas, Phys. Rev. Lett. 79, 341 (1997).
\item[{20.}] CDF and ${\rm DO\!\!\!\!/}$ collaborations: R. Culbertson,
Fermilab-Conf-97-277-E, Proc. SUSY97 (to be published).
\item[{21.}] M. Guchait and D.P. Roy, Phys. Rev. D52, 133 (1995).
\item[{22.}] H. Baer, C. Chen, F. Paige and X. Tata, Phys. Rev. D52,
2746 (1995). 
\end{enumerate}

\newpage
\begin{figure}[htbp]
\begin{center}
\epsfig{height=6cm,width=8cm,angle=0,file=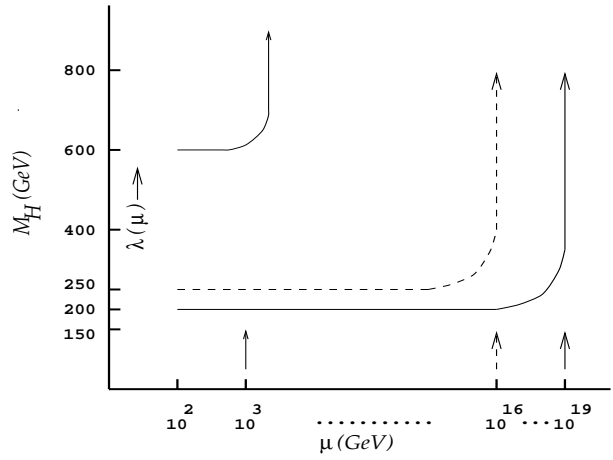}
\caption{The triviality bounds on the Higgs boson mass
corresponding to different cut-off scales, i.e. TeV, GUT and Planck
scales.} 
\label{fig:9808fig1}
\end{center}
\end{figure}
\begin{figure}[htbp]
\begin{center}
\epsfig{height=8cm,width=6cm,angle=90,file=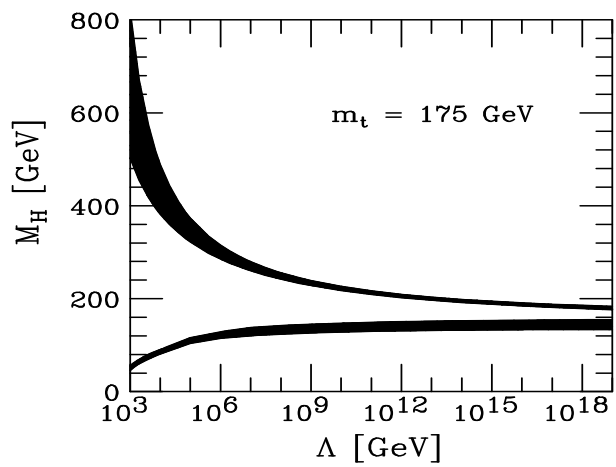}
\caption{The upper and lower bounds on the mass of the SM
Higgs boson as functions of the cutoff scale [4].}
\label{fig:9808fig2}
\end{center}
\end{figure}

\newpage

\begin{figure}[htbp]
\begin{center}
\epsfig{height=8cm,width=6cm,angle=-90,file=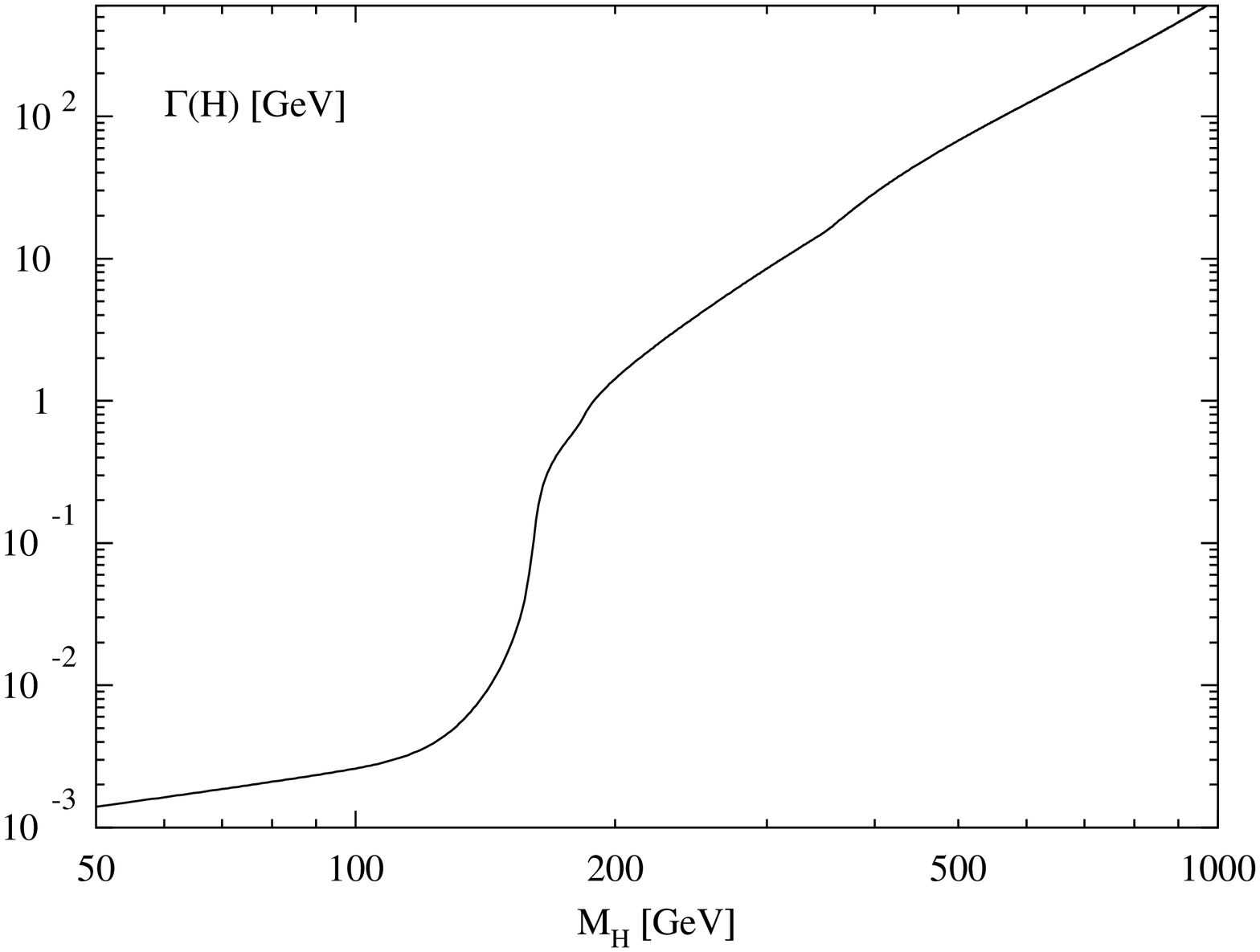}
\end{center}
\begin{center}
\epsfig{height=8cm,width=6cm,angle=-90,file=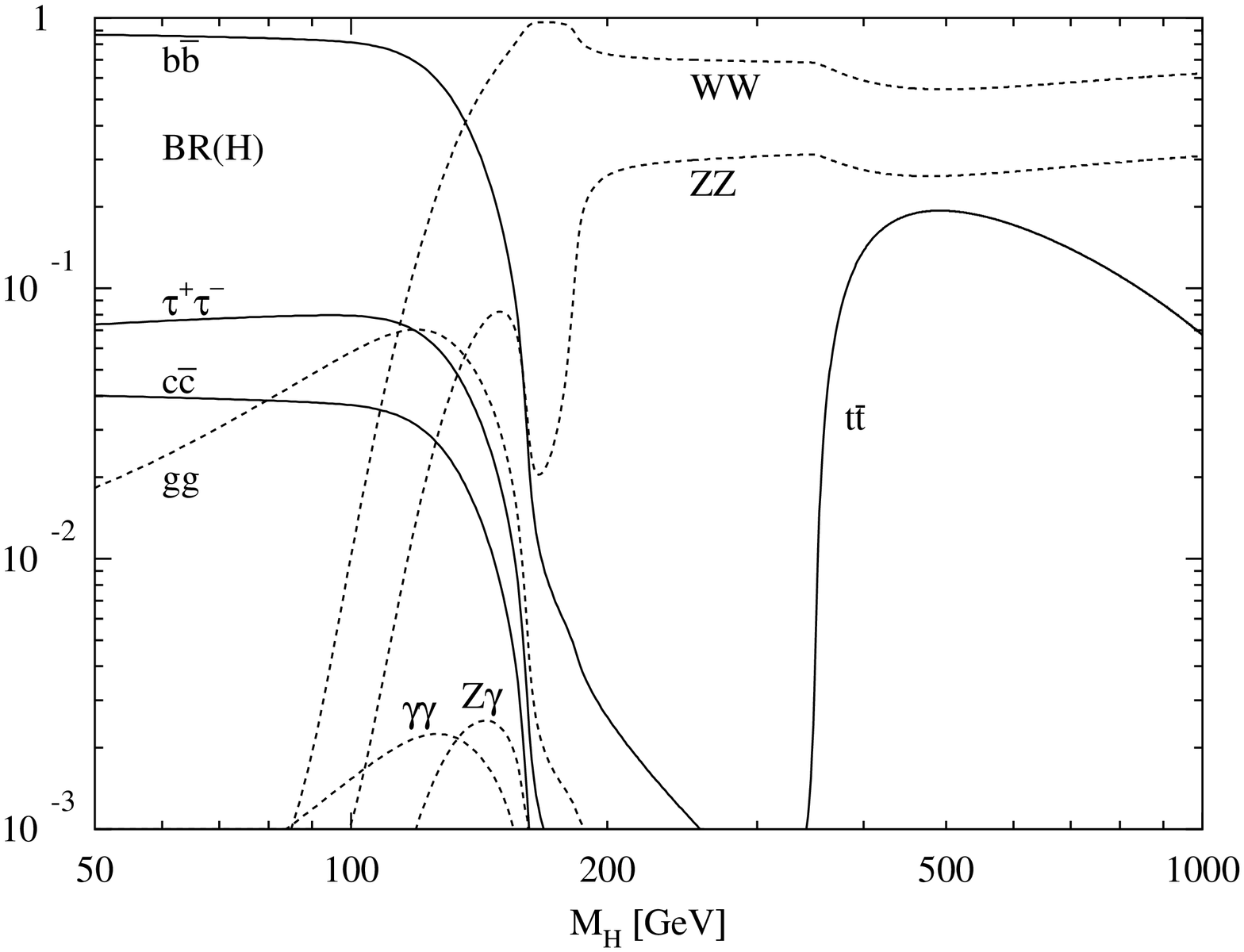}
\caption{Total decay width and the main branching ratios of
the SM Higgs boson [8].}
\end{center}
\end{figure}

\newpage
\begin{figure}[htbp]
\begin{center}
\epsfig{height=8cm,width=6cm,angle=-90,file=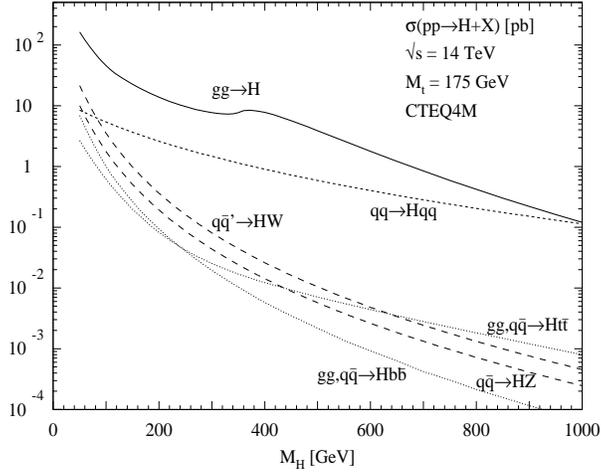}
\caption{Production cross-sections of the SM Higgs boson at LHC [8].}
\end{center}
\end{figure}
\begin{figure}[htbp]
\begin{center}
\epsfig{height=6cm,width=8cm,angle=0,file=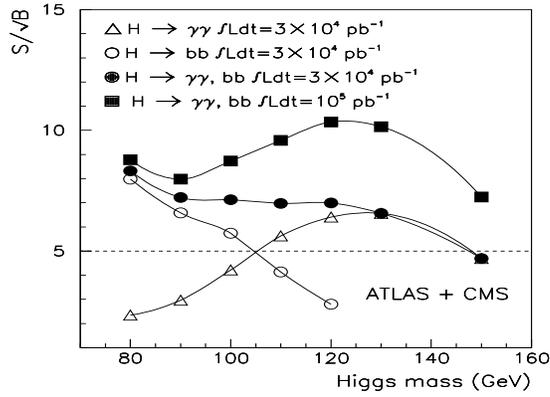}
\caption{The expected significance level of the SM Higgs signal at LHC
over the intermediate mass region [9].}
\end{center}
\end{figure}

\newpage
\begin{figure}[htbp]
\begin{center}
\epsfig{height=8cm,width=6cm,angle=-90,file=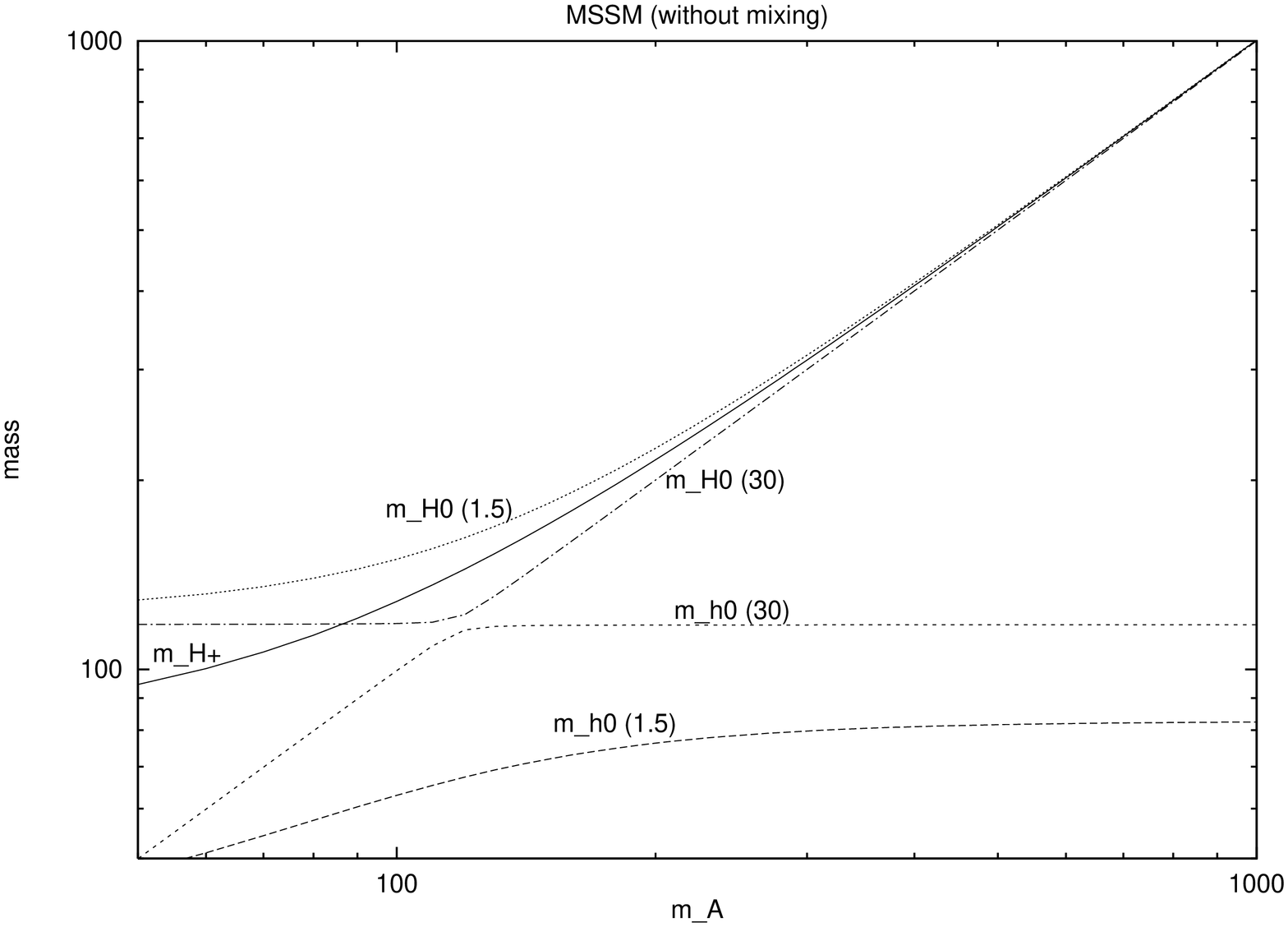}
\end{center}
\begin{center}
\epsfig{height=8cm, width=6cm, angle=-90, file=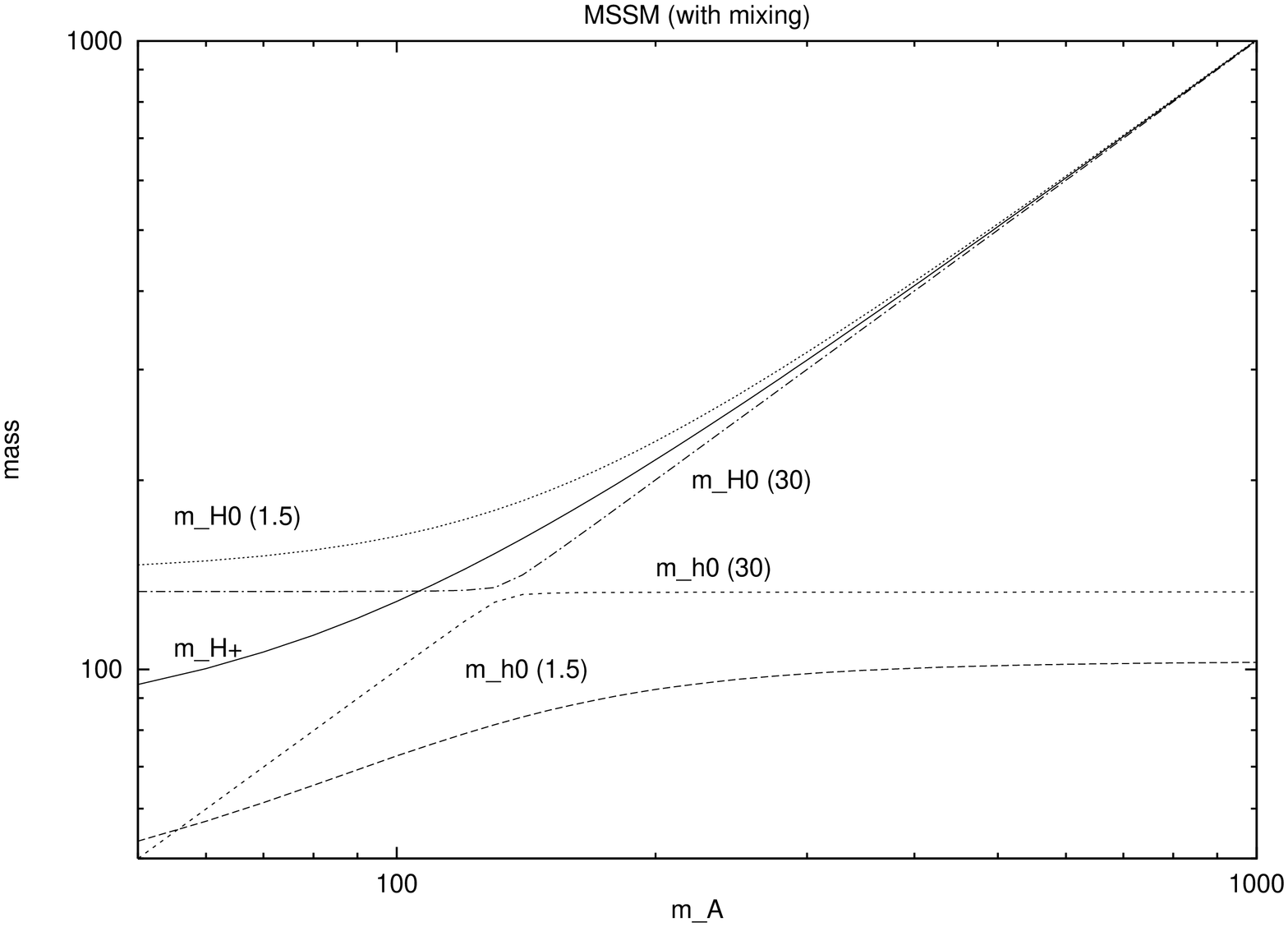}
\caption{Masses of the MSSM Higgs boson $h^0$,$H^0$ and $H^\pm$ as
functions of the pseudoscalar mass for $\tan\beta = 1.5$ and $30$. The
predictions without and with maximal stop mixing are shown
separately.}
\end{center}
\end{figure}

\newpage
\begin{figure}[htbp]
\begin{center}
\epsfig{height=14cm, width=14cm, angle=0, file=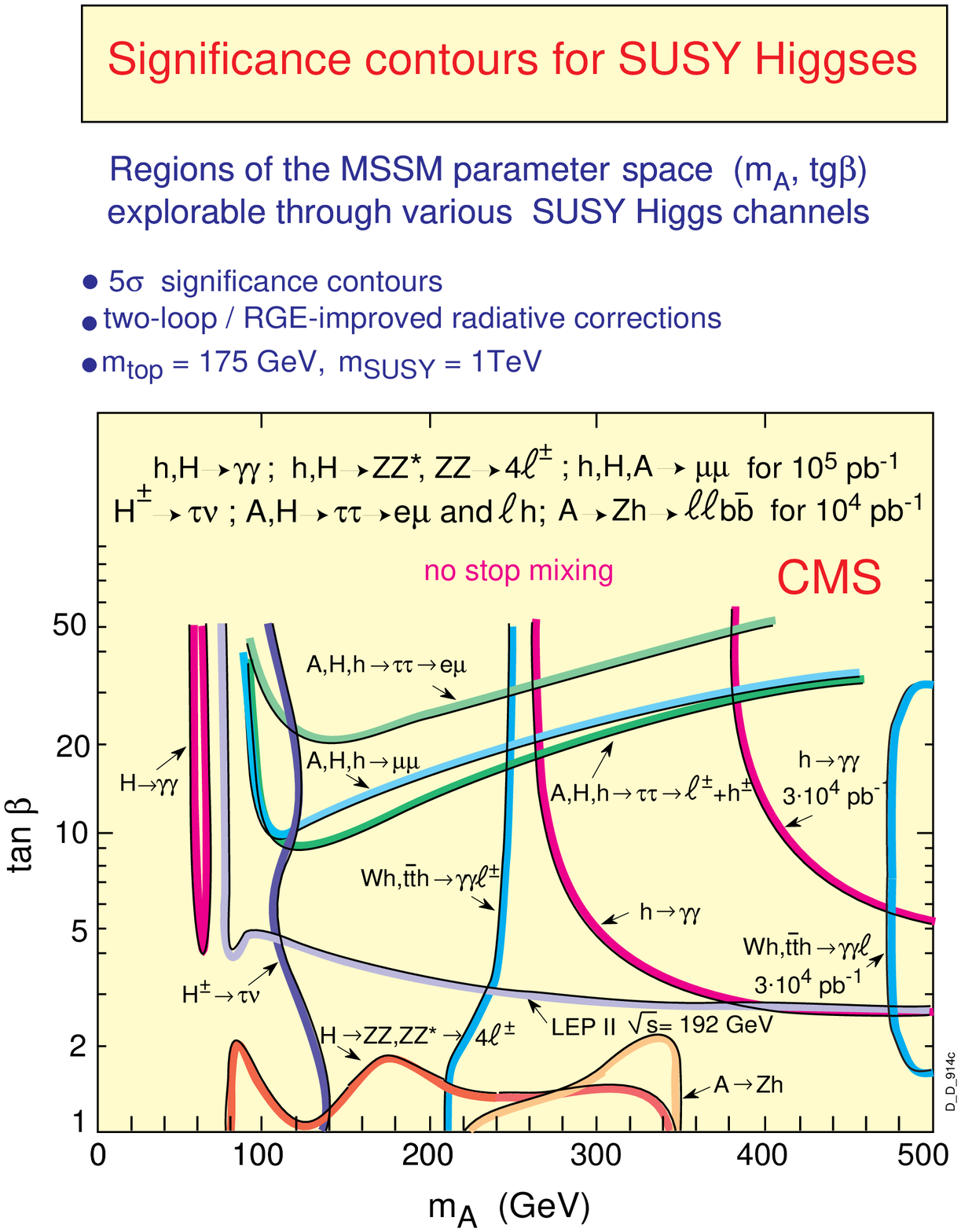}
\caption{Regions of the parameter space $(M_A - \tan\beta)$ covered by
the $5\sigma$ discovery contours of various MSSM Higgs signals from
the CMS experiment [13].}
\end{center}
\end{figure}

\newpage
\begin{figure}[htbp]
\begin{center}
\epsfig{height=10cm, width=10cm, angle=0, file=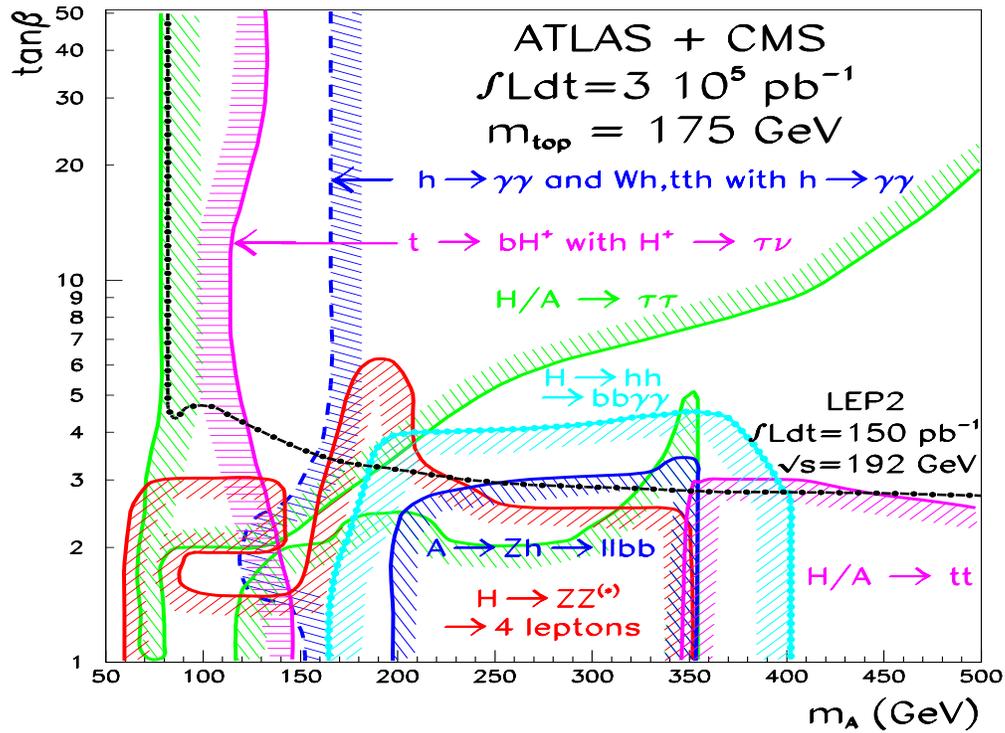}
\caption{Regions of the parameter space $(M_A - \tan\beta)$ covered by
the $5\sigma$ discovery contours of various MSSM Higgs signals from
the combined ATLAS + CMS experiments after 3 years of high luminosity
run of LHC [17].}
\end{center}
\end{figure}

\newpage
\begin{figure}[htbp]
\begin{center}
\epsfig{height=7cm, width=5cm, angle=90, file=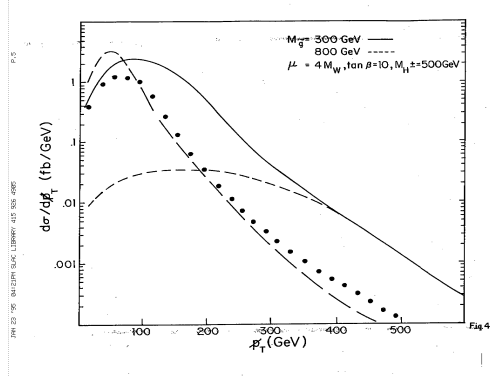}
\caption{The expected size of the LSD signals for 300 and 800 GeV
gluino production at LHC are shown against the accompanying
missing-$p_T$. The real and fake LSD backgrounds from $\bar t t$
production are shown by long dashed and dotted lines respectively
[21].}
\end{center}
\end{figure}
\begin{figure}[htbp]
\begin{center}
\epsfig{height=7cm, width=5cm, angle=90, file=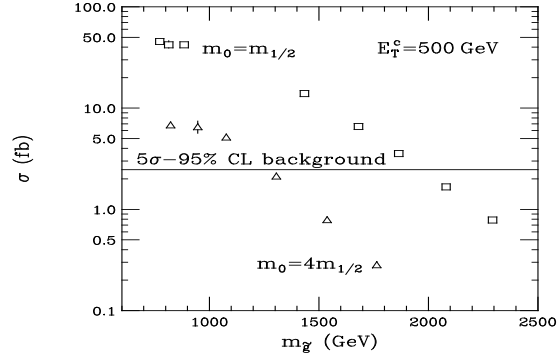}
\caption{The expected gluino signals at LHC from jets + missing-$E_T$
channel are shown for $M_{\tilde g} \simeq M_{\tilde q}$ (squares) and
$M_{\tilde g} \ll M_{\tilde q}$ (triangles). The $95\%$ CL background 
shown also corresponds to $5\sqrt{B}$ for the LHC luminosity 
of $10 fb^{-1}$ [22].}   
\end{center}
\end{figure}

\newpage
\begin{figure}[htbp]
\begin{center}
\epsfig{height=14cm, width=14cm, angle=0, file=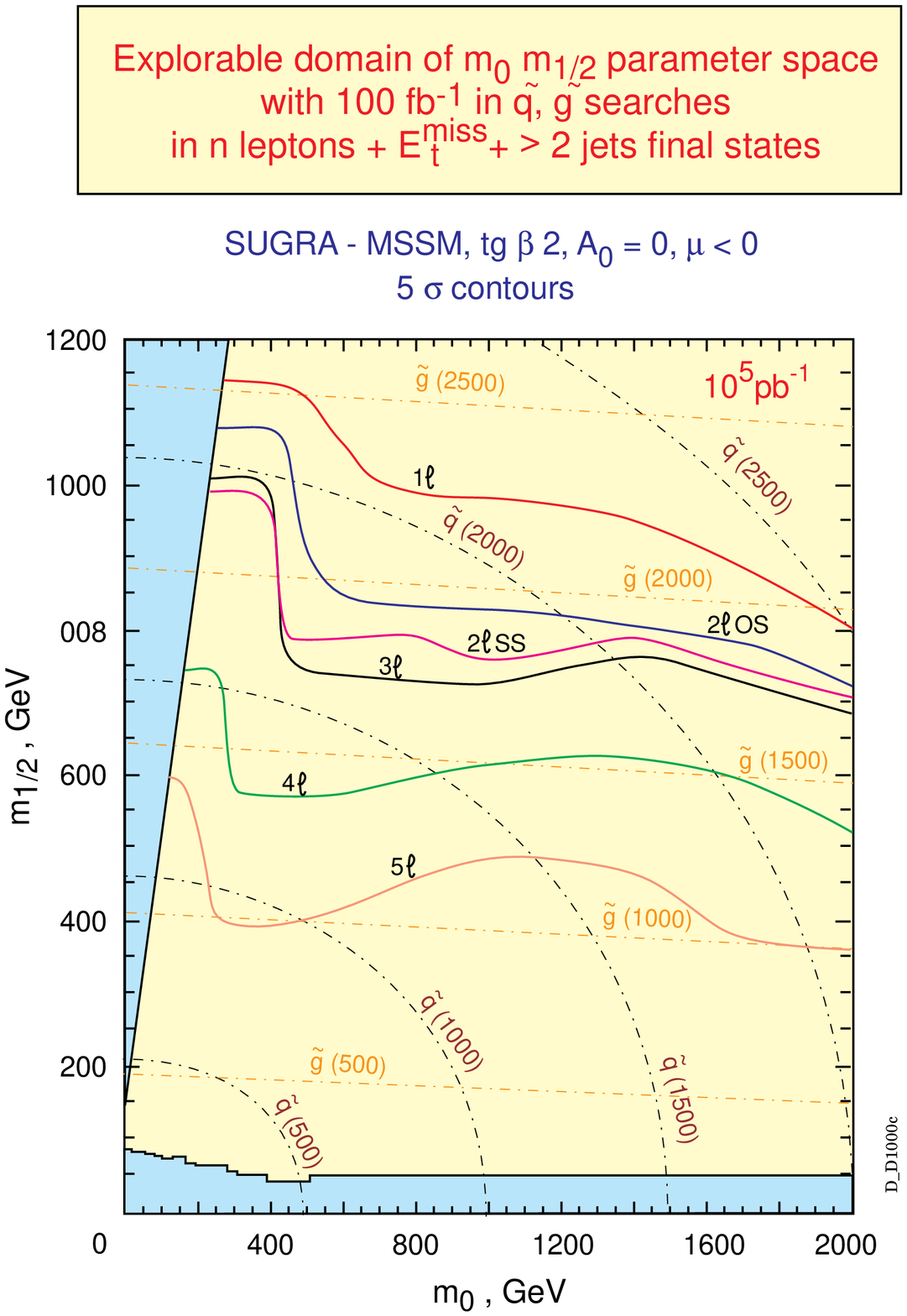}
\caption{The SUSY discovery limits of various leptonic channels at
LHC, where $2l$ $OS$ and $2l$ $SS$ denote opposite sign and same sign
dileptons [13].}
\end{center}
\end{figure}  

\end{document}